\documentclass[%
superscriptaddress,
preprint,
 amsmath,amssymb,
 aps,
pre,
]{revtex4-1}

\usepackage{graphicx}% Include figure files
\usepackage{dcolumn}% Align table columns on decimal point
\usepackage{bm}% bold math
\usepackage[colorlinks=true, linkcolor=blue, citecolor=blue]{hyperref}% add hypertext capabilities
\usepackage{amsmath}
\newcommand{\Oh}{\mathrm{Oh}}
\newcommand{\Real}{\mathrm{Re}}
\newcommand{\Imag}{\mathrm{Im}}

\begin{document}

%\preprint{APS/123-QED}

\title{Shape of a recoiling liquid filament}
\author{Francesco Paolo Cont\`o}
\affiliation{
School of Engineering and Materials Science, Queen Mary University of London, London E1 4NS, United Kingdom.
}
\author{Juan F. Mar\'in}
\affiliation{Departamento de F\'isica, Universidad de Santiago de Chile,
Av. Ecuador 3493, Estaci\'on Central, Santiago, Chile.}
\author{Arnaud Antkowiak}
\affiliation{
 Institut Jean Le Rond d'Alembert, UMR 7190 CNRS/UPMC, Sorbonne Universit\'es, F-75005 Paris, France.
}
\author{J. Rafael Castrej\'on Pita}
\affiliation{
School of Engineering and Materials Science, Queen Mary University of London, London E1 4NS, United Kingdom.
}
\author{Leonardo Gordillo\footnote{leonardo.gordillo@usach.cl}}
\affiliation{Departamento de F\'isica, Universidad de Santiago de Chile, Av. Ecuador 3493, Estaci\'on Central, Santiago, Chile.}
\email[]{leonardo.gordillo@usach.cl}

\date{\today}

\begin{abstract}
We study the capillary retraction of a Newtonian semi-infinite liquid
filament through analytical methods. We derive a long-time asymptotic-state expansion for the filament profile using a one-dimensional free-surface slender cylindrical flow model based on the three-dimensional axisymmetric Navier-Stokes equations. The analysis identifies three distinct length and time scale regions in the retraction domain: a steady filament section, a growing spherical blob, and an intermediate \textit{matching} zone. We show that liquid filaments naturally develop travelling capillary waves along their surface and a neck behind the blob. We analytically prove that the wavelength of the capillary waves is approximately 3.63 times the filament's radius at the inviscid limit. Additionally, the waves’ asymptotic wavelength, decay length, and the minimum neck size are analysed in terms of the Ohnesorge number. Finally, our findings are compared with previous results from the literature and numerical simulations in Basilisk obtaining a good agreement. This analysis provides a full picture of the recoiling process going beyond the classic result of the velocity of retraction found by Taylor and Culick. 
\end{abstract}

%\keywords{Drops, Breakup/Coalescence, Liquid bridges, Capillary waves}%Use showkeys class option if keyword
				%display desired
                              
\maketitle

%\tableofcontents

\section{Introduction\label{sec:introduction}}

\begin{figure}
\begin{centering}
\includegraphics[width=17cm]{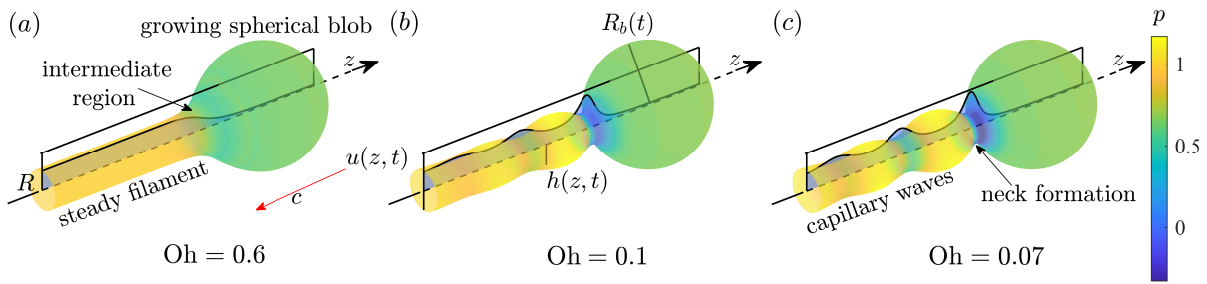}
\par\end{centering}
\caption{Recoiling of filaments at $t=33.8$ for ($a$) $\Oh = 0.6$,
($b$) $\Oh = 0.1$, and ($c$) $\Oh=0.07$. The time $t$ is scaled by the capillary time $t_{c}=\sqrt{\rho R^{3}/\sigma}$. Asymptotic solutions were obtained through a lubrication model. Velocity profiles (black solid lines) and pressure $p$ (colour scheme) are also displayed. \label{fig:recoil3D}}
\end{figure}

The capillary retraction of liquid filaments in a passive ambient fluid is extremely important in several industrial applications, such as atomisation, spraying, inkjet printing, and microfluidics \cite{basaran2002small}.
Additionally, their dynamics are a fundamental problem in nature. A liquid filament, free in the air, retracts by the action of surface tension. This is a classical problem that has been widely studied in the literature through theoretical, numerical and experimental methods \cite{keller1983breaking, stone1986experimental, stone1989relaxation,  schulkes1996contraction,  notz2004dynamics, Lohse2013, hoepffner2013recoil, castrejon2015plethora, wang2019fate}.
However, the first significant studies on this topic were performed by Taylor in 1959 \cite{taylor1959dynamics} and Culick in 1960 \cite{culick1960comments}, who focused on the capillary retraction of a thin planar fluid sheet. Following basic physics principles, i.e. mass and momentum conservation, they found that the fluid free edge accumulates mass from the film as it retracts, and a growing rim with a circular section is formed. The rim recedes at a retraction speed that tends to a constant value, i.e. the \textit{Taylor-Culick speed}, and is independent of the fluid viscosity as only inertial and capillary effects are considered. In the case of a cylindrical filament of radius $R$, the liquid is collected at the receding tip thus forming a growing spherical blob as depicted in Fig.~\ref{fig:recoil3D}. Mass and momentum balance applied to a domain enclosing the blob yields the Taylor-Culick speed,

\begin{equation}
c=\sqrt{\frac{\sigma}{\rho R}},\label{eq. T-C_speed}
\end{equation}
where $\sigma$ is the surface tension and $\rho$ is the liquid density. Subsequent studies by Savva \& Bush \cite{savva2009viscous} showed that the retraction velocity does converge to the Taylor-Culick value after an initial unsteady state, confirming its validity as an asymptotic limit. Indeed, such a powerful prediction is derived from very simple arguments: mass and momentum balance applied at a global scale (net force acting on the growing blob). In addition, Savva \& Bush in 2009 \cite{savva2009viscous} and  Brenner \& Gueyffier in 1999 \cite{brenner1999bursting} showed that the liquid viscosity, $\mu$, determines the transient-state characteristic time, the interface stability, and the filament profile. It is remarkable that the problem is governed by a single dimensionless parameter, the Ohnesorge number  $\Oh\equiv\mu/\sqrt{\rho\sigma R}$, which represents the ratio of viscous and capillary forces. Examples of the behaviour of the filament retraction for different $\Oh$ are shown in Fig.~\ref{fig:recoil3D}.

In this report, we study the filament interface evolution in the long-time limit, i.e. when the blob radius is much larger than the filament radius, using a local asymptotic analysis. In Section $\S$\ref{sec:filament_equations}, we present a set of two one-dimensional equations for a free surface slender filament liquid under the lubrication approximation. We use this model to analyse the filament dynamics in its quasi-steady state. We show that the long-time asymptotic solution is separated into three regions that have different length and time scales: the steady cylinder far from the blob, the receding blob, and an intermediate matching zone (see Fig.~\ref{fig:recoil3D}). In Section~$\S$\ref{sec:solutions_filament}, the three different solutions are ``stitched'' together to provide a unique asymptotic matching solution of the evolving filament shape. Travelling capillary ripples and the neck connecting the blob and the filament, which can be seen under some conditions, are analysed in terms of the Ohnesorge number. These results are important as they provide a full and formal description of the Taylor-Culick regime.

\section{Filament equations and local solutions\label{sec:filament_equations}}

Here, we consider an incompressible axisymmetric liquid filament of viscosity $\mu$ and density $\rho$ surrounded by an inert ambient gas with negligible density. The boundaries of the liquid are taken as free surfaces with surface tension $\sigma$. The filament is a semi-infinite cylinder of radius $R$ aligned to the $z$-axis and
its free rim recoils as time $t$ evolves.

Following the approach of Eggers \& Dupont \cite{eggers1994drop}, the system dynamics can be described through a lubrication model. In this long-wavelength limit, the problem is reduced to a set of two coupled equations along the axial direction $z$ for two scalar fields: the filament local radius $h(z,t)$ and the mean axial velocity $u(z,t)$. The two equations correspond to local mass conservation and local momentum balance. In dimensionless form ($h,z\sim R$, $u\sim c$ and $t\sim R/c$), these are:
\begin{align}
\partial_{t}\left(h^{2}\right)+\partial_{z}\left(h^{2}u\right) & =0,\label{eq. lubri_1_nd}\\
\partial_{t}\left(h^{2}u\right)+\partial_{z}\left(h^{2}u^{2}\right) & =\partial_{z}\left(3\Oh\,h^{2}\partial_{z}u+h\left[\frac{1+\left(\partial_{z}h\right){}^{2}+h\partial_{zz}h}{\sqrt{1+\left(\partial_{z}h\right){}^{2}}^{3}}\right]\right).\label{eq. lubri_2_nd}
\end{align}
Gravity has been neglected and the full non-linear expression for the capillary term has been included as suggested by Eggers \cite{PhysRevLett.71.3458}.

Under the action of the surface tension, the filament contracts at a rate that asymptotically reaches the Taylor-Culick velocity ($c=-1$ in dimensionless form). For convenience, a reference frame moving at the Taylor-Culick velocity is chosen. The boundary condition for Eqns. \eqref{eq. lubri_1_nd} and \eqref{eq. lubri_2_nd} at $z=-\infty$ is hence $h=1$ and $u=1$, representing the unperturbed filament radius far away from the tip, and the influx velocity at the blob. Likewise, at the tip $z=z_{T}\left(t\right)$, the boundary condition is $h=0$, $\partial_{z}h=-\infty$ and $u=\partial_{t}z_{T}$, which sets the symmetry and the kinematic condition at the receding tip. In the next sections, we derive a long-time asymptotic expression for the filament profile from Eqns. \eqref{eq. lubri_1_nd} and \eqref{eq. lubri_2_nd}. 

\subsection{Steady filament\label{subsec:far_field}}

\begin{figure}
\begin{centering}
\includegraphics[width=\textwidth]{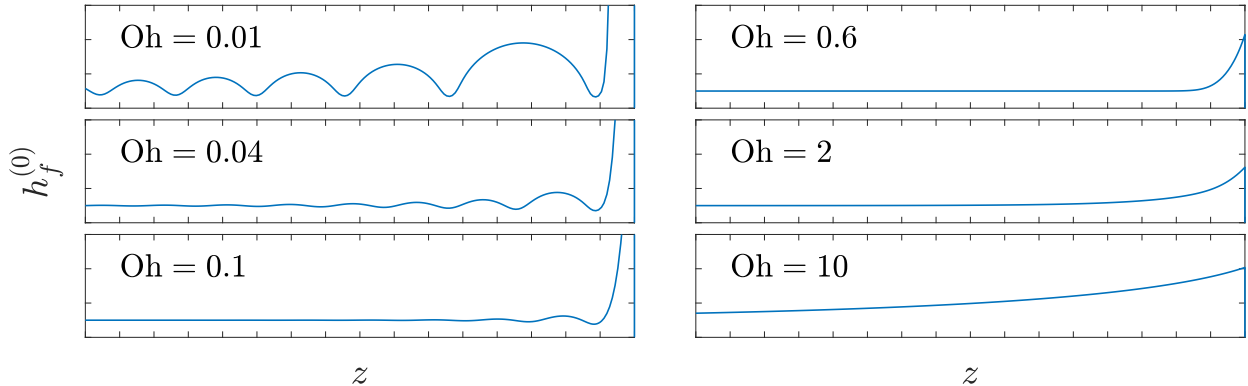}
\par\end{centering}
%\caption{Numerical solutions for the profile of the steady filament region (via Eqn. \ref{eq. ode}, for $z_{0}=0$) for different Ohnesorge numbers. The distance between ticks is of $2R$. As the $\Oh$ number grows, ripples decay faster. As explained in Section $\S$\ref{subsec:capillaryripples_decaylength}, these ripples, or otherwise called capillary waves, appear for $\Oh \leq 3^{-\frac{1}{2}}$. A decay length\label{fig:separatrix_envelopes}}
\caption{Numerical solutions for the profile of the steady filament region (via Eqn. \eqref{eq. ode} with $z_{0}=0$) for different Ohnesorge numbers. The distance between ticks is $2R$. As $\Oh$ grows, solutions show slow decaying ripples ($\Oh=0.01$), fast decaying ripples ($\Oh=0.1$), fast smooth decay ($\Oh=0.6$) and slow smooth decay ($\Oh=10.0$). As explained in $\S$\ref{subsec:capillaryripples_decaylength}, oscillations appear for $\Oh<\Oh^{*}=3^{-\frac{1}{2}}$. At the $z\rightarrow\infty$ limit, the solution shows the exponential behaviour of Eqn.~\eqref{eq. separatrix_expansion}, with steeper growths for smaller $\Oh$ numbers. \label{fig:separatrix_envelopes}}
\end{figure}

Here, we seek solutions for the steady region far from the blob. The following expressions for the filament dimensionless radius and the axial velocity are proposed:
\begin{eqnarray}
h_{f}\left(z,t\right) & = & h_{f}^{\left(0\right)}\left(z\right)+h_{f}^{\left(1\right)}\left(z,t\right)+\ldots,\\
u_{f}\left(z,t\right) & = & u_{f}^{\left(0\right)}(z)+u_{f}^{\left(1\right)}\left(z,t\right)+\ldots,
\end{eqnarray}
where $h_{f}^{\left(1\right)}\left(z,t\right)$ and $u_{f}^{\left(1\right)}\left(z,t\right)$ are higher-order corrections, such that $h_{f}^{\left(1\right)}/h_{f}^{\left(0\right)}$, $h_{f}^{\left(2\right)}/h_{f}^{\left(0\right)}$, ... etc, and $u_{f}^{\left(1\right)}/u_{f}^{\left(0\right)}$, $u_{f}^{\left(2\right)}/u_{f}^{\left(0\right)}$, ... etc, vanish as $t\to\infty$. By replacing these expressions into Eqns. \eqref{eq. lubri_1_nd} and \eqref{eq. lubri_2_nd}, we obtain, at the leading order:
\begin{eqnarray}
\partial_{z}\left(h^{2}u\right) & = & 0,\label{eq. mass_farfield}\\
\partial_{z}\left(h^{2}u^{2}\right) & = & \partial_{z}\left(3\Oh\,h^{2}\partial_{z}u+h\left[\frac{1+\left(\partial_{z}h\right){}^{2}+h\partial_{zz}h}{\sqrt{1+\left(\partial_{z}h\right){}^{2}}^{3}}\right]\right), \label{eq. momentum_farfield}
\end{eqnarray}
where $h_{f}^{(0)}(z)$ and $u_{f}^{(0)}(z)$ have been written as $h$ and $u$ to simplify the notation. The axial velocity $u$ can be solved in terms of $h$ from Eqn. \eqref{eq. mass_farfield} using the boundary conditions, and then combined with Eqn. \eqref{eq. momentum_farfield} to obtain the second-order ordinary differential equation
\begin{equation}
\frac{1+6\Oh\cdot h\partial_{z}h}{h^{3}}=\frac{1+\left(\partial_{z}h\right){}^{2}+h\partial_{zz}h}{\sqrt{1+\left(\partial_{z}h\right){}^{2}}^{3}}.\label{eq. ode}
\end{equation}
This equation can only be integrated numerically, but several of its properties can be studied analytically. First, as expected, it is invariant under translations. Second, it can be shown that all solutions diverge for a finite $z$ except for one, which is the one to be matched to a growing blob. The analytic expansion around $z\rightarrow\infty$ is
\begin{equation}
\lim_{z\rightarrow\infty}h_{f}^{(0)}(z)=\mathrm{e}^{\frac{z-z_{0}}{\sqrt{3\Oh}}}+\mathcal{O}\left(\mathrm{e}^{-\frac{z-z_{0}}{\sqrt{3\Oh}}}\right),\label{eq. separatrix_expansion}
\end{equation}
where $z_{0}$ is a translational invariance constant. A set of these solutions, for different Ohnesorge numbers, is shown in Fig.~\ref{fig:separatrix_envelopes}. Depending on the $\Oh$ value, this region is either featured by decaying ripples (capillary waves) or a smooth spatial decay. These features are further analysed
in Section $\S$\ref{subsec:capillaryripples_decaylength}.

\subsection{The Growing Blob\label{subsec:rim}}

The blob region is characterised by two quantities: a time-dependent length scale, its radius $R_{b}\left(t\right)$, and the constant liquid mass influx as the filament recedes, $h^{2}u=1$.
Consequently, the following self-similar solutions are proposed:
\begin{eqnarray}
h_{b}\left(z,t\right) & = & R_{b}\left[h_{b}^{\left(0\right)}\left(\zeta\right)+h_{b}^{\left(1\right)}\left(\zeta,t\right)+\ldots\right],\\
u_{b}\left(z,t\right) & = & R_{b}^{-2}\left[u_{b}^{\left(0\right)}\left(\zeta\right)+u_{b}^{\left(1\right)}\left(\zeta,t\right)+\ldots\right],
\end{eqnarray}
where $\zeta\equiv R_{b}^{-1}z$, and $h_{b}^{\left(1\right)}\left(\zeta,t\right)$ and $u_{b}^{\left(1\right)}\left(\zeta,t\right)$ are higher-order corrections in $t$. Combining these solutions with Eqns. \eqref{eq. lubri_1_nd} and \eqref{eq. lubri_2_nd}, the following leading-order relationships are obtained:
\begin{eqnarray}
2R_{b}^{2}\dot{R}_{b}h^{2}-2R_{b}^{2}\dot{R}_{b}\zeta h\partial_{\zeta}h+\partial_{\zeta}\left(h^{2}u\right) & = & 0,\label{eq. rim_mass}\\
\partial_{\zeta}\left(\frac{1+\left(\partial_{\zeta}h\right){}^{2}+h\partial_{\zeta\zeta}h}{\sqrt{1+\left(\partial_{\zeta}h\right){}^{2}}^{3}}\right) & = & 0,\label{eq. rim_momentum}
\end{eqnarray}
where $\dot{R}_{b}\equiv\partial_{t}R_{b}$, $h\equiv h_{b}^{\left(0\right)}\left(\zeta\right)$ and $u\equiv u_{b}^{\left(0\right)}\left(\zeta\right)$. The solution for Eqn. \eqref{eq. rim_momentum}, which is uncoupled from Eqn. \eqref{eq. rim_mass}, can be easily obtained, and is given by the semicircle
\begin{equation}
h\left(\zeta\right)=\sqrt{2\zeta-\zeta{}^{2}.}\label{eq. circle}
\end{equation}
This solution has been chosen so that the left tip of the semicircle coincides with the origin. Replacing this solution in Eqn. \eqref{eq. rim_mass}, and forcing the flux $h^{2}u$ at the tip ($\zeta=2$) to be zero, we obtain a condition for the time-dependent length scale $R_{b}\left(t\right)$, i.e. $R_{b}^{2}\dot{R}_{b}=1/4$. Importantly, this condition is identical to the classical Taylor-Culick argument that balances the growth of the blob radius with the filament collected volume:
$\partial_{t}\left(4\pi R_{b}^{3}/3\right)=\pi$. Hence, in the asymptotic limit, the radius of the blob grows as $R_{b}\left(t\right) \rightarrow \left(3t/4\right)^{1/3}.$
The blob shape, in terms of the variables $z$ and $t$, is given by
\begin{equation}
h_b^{(0)}\left(z,t\right)=\sqrt{2R_{b}\left(t\right)z-z{}^{2}},
\end{equation}
where $h_b^{(0)}(z,t)\equiv R_b h_b^0(\zeta)$ and its expansion around $z\rightarrow0$ is
\begin{equation}
\lim_{z\rightarrow0^{+}}h_{b}^{(0)}\left(z,t\right)=6^{1/6}t^{1/6}z^{1/2}+\mathcal{O}\left(t^{-1/6}z^{3/2}\right).\label{eq. rim_expansion}
\end{equation}
Another important observation is that the blob, in the asymptotic limit, is completely
independent from viscosity, as evidenced by the momentum Eqn. \eqref{eq. rim_momentum}. Physically, the enlargement of the spatial scales in the growing blob slows the flow into a quasi-steady state that leads to the rise of surface tension as the unique dominant force.

\subsection{Intermediate matching zone\label{subsec:intermediate_region}}
An intermediate region is required to match the zeroth-order solutions of the steady filament and the growing blob obtained in the previous sections. At the leading order, the profile of this intermediate zone has to fit both the steady-filament solution ($h\propto\exp\left[z/\sqrt{3\Oh}\right]$) for $z\rightarrow-\infty$, and the blob ($h\propto t^{1/6}z^{1/2}$) for $z\rightarrow\infty$. Provided that the flux $h^{2}u=1$ is conserved across the region, we propose the following ansatz:
\begin{eqnarray}
h_{m}\left(z,t\right) & = & t^{1/6}\left[h_{m}^{\left(0\right)}\left(\xi \right)+h_{m}^{\left(1\right)}\left(\xi,t\right)+\ldots\right],\\
u_{m}\left(z,t\right) & = & t^{-1/3}\left[u_{m}^{\left(0\right)}\left(\xi \right)+u_{m}^{\left(1\right)}\left(\xi,t\right)+\ldots\right],
\end{eqnarray}
where $\xi\equiv z$ is merely a spatial variable, and $h_{m}^{\left(1\right)}\left(\xi,t\right)$ and $u_{m}^{\left(1\right)}\left(\xi,t\right)$ are higher-order corrections. Combining these expressions with Eqns. \eqref{eq. lubri_1_nd} and \eqref{eq. lubri_2_nd}, we obtain at the leading order:
\begin{eqnarray}
\partial_{\xi}\left(h^{2}u\right) & = & 0,\label{eq. intermediate_mass}\\
\partial_{\xi}\left[h\left(\frac{\left(\partial_{\xi}h{}^{2}\right)+h\partial_{\xi\xi}h}{\left(\partial_{\xi}h\right){}^{3}}\right)\right]+3\Oh\partial_{\xi}\left(h^{2}\partial_{\xi}u\right) & = & 0,\label{eq. intermediate_momentum}
\end{eqnarray}
where $h\equiv h_{m}^{\left(0\right)}\left(\xi\right)$ and $u\equiv u_{m}^{\left(0\right)}\left(\xi\right)$.
Eqn. \eqref{eq. intermediate_mass} can easily be integrated by taking the flux boundary condition $h^{2}u\mid_{\xi=-\infty}=1$.
This yields $u=h^{-2}$, which can then be plugged into Eqn. \eqref{eq. intermediate_momentum} to obtain the following equation
\begin{equation}
6\Oh\left(\partial_{\xi}h\right)^{4}=h^{2}\left[(\partial_{\xi}h)^{2}+h\partial_{\xi\xi}h\right].\label{eq. ode_intermediate}
\end{equation}
This ordinary differential equation is now integrated imposing a limiting condition that matches the growing blob region, i.e. $h_{m}^{\left(0\right)}\propto\xi^{1/2}$ at $\xi\rightarrow\infty$. The result is then integrated once again, yielding 
\begin{equation}
h\left(\xi\right)=\alpha\left(\frac{3}{4}\Oh\right)^{1/4}H\left(\Xi\equiv\frac{2\xi}{\sqrt{3\Oh}}\right),\label{eq. intermediate_implicit}
\end{equation}
where $\alpha$ is a constant and $H$ is given implicitly by 
\[
\Xi=\sqrt{H^{4}+1}-\arctan\left(\frac{1}{\sqrt{H^{4}+1}}\right).
\]
Expanding Eqn. \eqref{eq. intermediate_implicit} into its two corresponding limits, we obtain
\begin{eqnarray}
\lim_{z\rightarrow-\infty}h_{m}^{(0)}\left(z,t\right) & = & qt^{1/6}\mathrm{e}^{\frac{z}{\sqrt{3\Oh}}}+\mathcal{O}\left(t^{1/6}\mathrm{e}^{\frac{5z}{\sqrt{3\Oh}}}\right),\label{eq. intermediate_leftlimit}\\
\lim_{z\rightarrow+\infty}h_{m}^{(0)}\left(z,t\right) & = & \alpha t^{1/6}z^{1/2}+\mathcal{O}\left(t^{1/6}z^{-3/2}\right),\label{eq. intermediate_rightlimit}
\end{eqnarray}
where $h_m^{(0)}(z,t)\equiv t^{1/6} h_m^0(\xi)$ and $q\equiv\alpha\left(3\mathrm{e}^{-2}\Oh\right)^{1/4}$. Equations \eqref{eq. intermediate_leftlimit} and \eqref{eq. intermediate_rightlimit} have already been written in terms of the original physical variables.
It is important to point out that the intermediate matching region arises from the balance between viscous and capillary forces as shown by Eqn. \eqref{eq. intermediate_momentum}. Nonetheless, as the filament becomes thicker, the capillary force overcomes viscosity leading to a viscosity-independent behaviour, suitable to be matched with the growing blob.

\begin{figure}
\centering{}\includegraphics{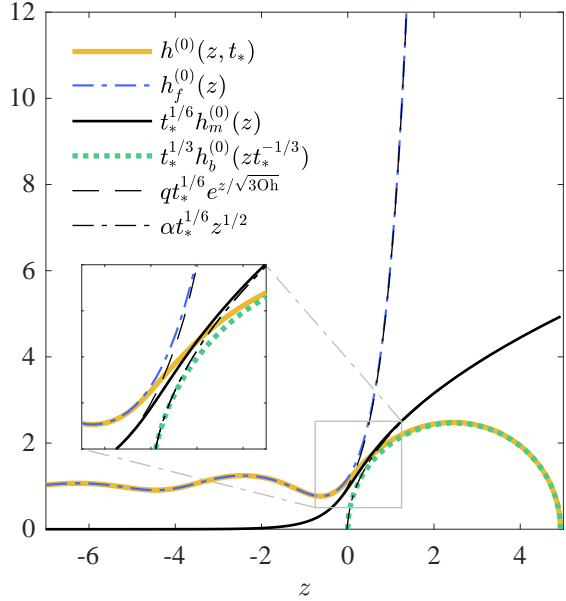} \caption{Asymptotic solutions for the asymptotic matching of the steady filament, the intermediate and circular blob regions for $\Oh=0.1$ and $t=t_*=20$ in dimensionless units. The asymptotic tails beyond the matched regions are also shown. \label{fig:matched_solution}}
\end{figure}

%% Figures 1 and 2 can be compressed in a single figure

\section{The evolving shape of a filament\label{sec:solutions_filament}}

\begin{figure}
\begin{centering}
\includegraphics[width=\textwidth]{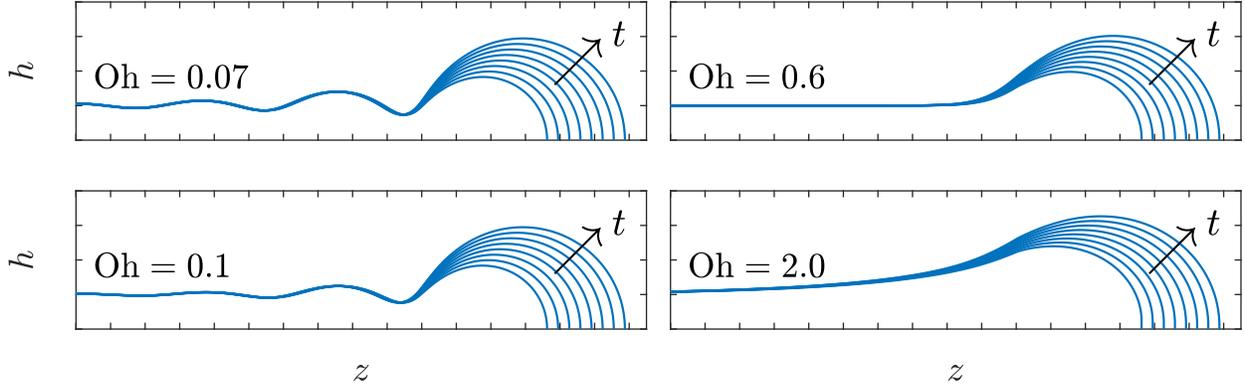}
\par\end{centering}
\caption{Filament profiles at $t=7.94$, $t = 10.25$, $t = 12.96$, $t = 16.12$, $t = 19.75$, $t = 23.88$, $t = 28.56$ and $t = 33.81$ in dimensionless units for $\Oh=0.07,\Oh = 0.10, \Oh = 0.60$ and $\Oh = 2.00$ in a Taylor-Culick velocity frame of reference. Curves were obtained following the asymptotic approach of Eqn. \eqref{eq. global_solution}. Here, the tick distance is equal to $R$. \label{fig:rim_animation}}
\end{figure}

The matching of the intermediate region solution, through Eqns. \eqref{eq. intermediate_leftlimit} and \eqref{eq. intermediate_rightlimit}, with the far-field and blob solutions, via Eqns. \eqref{eq. separatrix_expansion} and \eqref{eq. rim_expansion}, sets the coefficients $\alpha=6^{1/6}$ and $z_{0}=-\sqrt{3\Oh}\ln\left(qt^{1/6}\right)$.
Notice that the slow dependence of $z_{0}$ on time only modifies higher-order equations for $h_{f}^{\left(1\right)}$, leaving Eqns. \eqref{eq. mass_farfield} and \eqref{eq. momentum_farfield}
unaffected. Hence, the leading order solution for the filament profile that satisfies the lubrication Eqns. \eqref{eq. lubri_1_nd} and \eqref{eq. lubri_2_nd} with the proper boundary conditions is
\begin{equation}
h^{\left(0\right)}\left(z,t\right)  =  h_{f}^{\left(0\right)}\left(z\right)+t^{1/6}h_{m}^{\left(0\right)}\left(z\right)+t^{1/3}h_{b}^{\left(0\right)}\left(zt^{-1/3}\right)-qt^{1/6}\mathrm{e}^{\frac{z}{\sqrt{3\Oh}}}-\alpha t^{1/6}z^{1/2}.\label{eq. global_solution}
\end{equation}
An example of the leading order solution for a given time, including local solutions and asymptotic matched limits, is shown in Fig.~\ref{fig:matched_solution}. Evolving solutions based on Eqn. \eqref{eq. global_solution} are depicted in Fig.~\ref{fig:rim_animation}. Leading order expressions for $u\left(z,t\right)$ can be obtained under the same framework, and the pressure inside the filament can be directly derived via the formalism of Eggers \& Dupont \cite{eggers1994drop}. The filament pressure $p$ is shown as the colour scheme in Fig.~\ref{fig:recoil3D}. Additionally, higher-order corrections can be calculated too, following the scheme for the 2D case presented by Gordillo \emph{et al.} in 2011 \cite{gordillo2011asymptotic}. 

\subsection{Capillary waves and the decay length\label{subsec:capillaryripples_decaylength}}

The recoiling filament solution features travelling capillary waves that escape from the growing blob at the Taylor-Culick velocity.
In the chosen frame of reference, the capillary waves are steady and hence can be approximated by  $h_{f}^{\left(0\right)}\left(z\right)\approx1+\epsilon e^{-\mathrm{i}kz}$, with $\epsilon\ll1$. Combining this with Eqn. \eqref{eq. ode} and linearising, leads to the capillary wavenumber
\begin{equation}
k=3\mathrm{i}\Oh\pm\sqrt{3-9\Oh^{2}},
\label{eq: wavenumber}
\end{equation}
whose values undergo a transition between complex and pure imaginary values at the threshold $\Oh^{*}=3^{-\frac{1}{2}}\approx0.577.$ For $\Oh\geq\Oh^{*}$, the wavenumber $k$ is strictly imaginary and thus an exponential decay is observed as $z\rightarrow-\infty$. In contrast, for $\Oh<\Oh^{*}$, $k$ has a real component and the filament develops the well-known capillary waves modulated by the exponential decay. The dimensionless wavelength of these capillary ripples is given by $\lambda=2\pi/\Real\left(k\right)$, which as $\Oh\rightarrow0$, takes the value of $\lambda_{Oh=0}=2\pi/\sqrt{3}\approx3.63$.
The linear analysis also reveals that the decay length, $\ell=1/\Imag\left(k\right)$, decays as $(3\Oh)^{-1}$ for low Ohnesorge numbers, and as $\left(2\Oh\right)$ in the large-$\Oh$ limit. Figure \ref{fig:lambda+hmin} shows the wavelength $\lambda$ and decay length $\ell$ as functions of $\Oh.$ 

We also studied the linear dispersion relation of the full Navier-Stokes equations from the work of Rayleigh \cite{rayleigh1892xvi} and looked for waves whose phase velocity, $\omega/k$, is equal to the Taylor-Culick velocity given in Eqn. \eqref{eq. T-C_speed}. Results are shown as dashed lines in Fig.~\ref{fig:lambda+hmin}. The qualitative agreement between the linearised lubrication approximation and full linearised Navier-Stokes equations is remarkably good, showing similar thresholds and asymptotic limits. Furthermore, the agreement is quantitatively excellent for $\Oh>1$.

\subsection{Neck thickness\label{subsec:neck_Oh}}

\begin{figure}
\centering{}\includegraphics{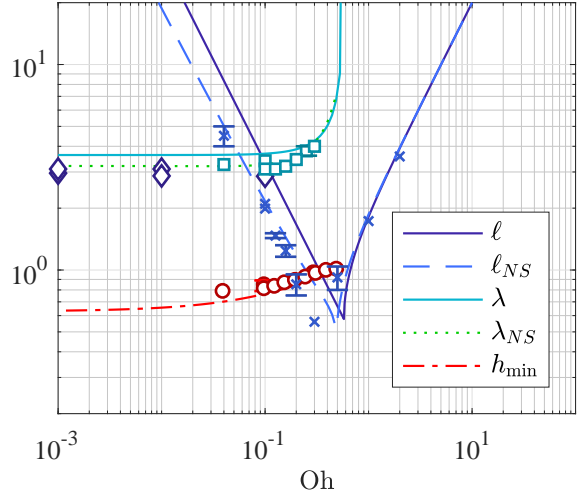}\caption{Wavelength, decay length, and neck thickness of the asymptotic solutions of lubrication equations as a function of $\Oh$. Additionally, the wavelength and decay length obtained from linearisation of full Navier-Stokes equations are shown as dashed lines. Numerical results are shown for comparison, wavelengths from Notz \& Basaran \cite{notz2004dynamics} ($\diamond$) and Basilisk ($\square$), decay lengths from Basilisk ($\times$), and neck thickness from Basilisk ($\circ$). Error bars are not shown when these are smaller than the symbol size. \label{fig:lambda+hmin}}
\end{figure}

A direct consequence of the existence of the capillary waves on the filament is the appearance of the \textit{neck}, i.e. a finite global minimum right behind the blob. Figure \ref{fig:lambda+hmin} shows the filament neck thickness $h_{\min}$ as a function of the Ohnesorge number. The results show that as $\Oh$ decreases, $h_{\min}$ converges to a finite value.
This value can be obtained by setting $\Oh=0$ in Eqn. \eqref{eq. ode}, and then integrating along $\mathit{z}$ using the boundary condition $h\rightarrow\infty$ as $z\rightarrow\infty$. The result yields
$\sqrt{1+\left(\partial_{z}h\right)^{2}}=4h^{3}$.
Setting $\partial_{z}h = 0$, we obtain an equation for the minimum neck, whose solution is
\begin{equation}
\lim_{\Oh\rightarrow0}h_{min}=4^{-\frac{1}{3}}\approx0.63,\label{eq. hmin_Oh=00003D0}
\end{equation}
which correctly matches the curve trend shown in Fig.~\ref{fig:lambda+hmin}. It is important to note that the asymptotic solution in this model does not display any critical Ohnesorge value at which $h_{\min}\rightarrow0$, which is contrary to what is seen in nature and numerical simulations.
Indeed, this suggests that filament pinch-off occurs through a dynamic instability.

\section{Validation\label{sec:validation}}

We analysed existing data in the literature for the purpose of comparisons with our findings. In the work of Notz \& Basaran \cite{notz2004dynamics}, simulations of long liquid filaments were performed at $\Oh =0.001$, $0.010$, $0.100$, and $1.000$. Capillary waves were seen for all the cases except for $\Oh = 1.000$, implying that the critical limit is somewhere in the range of $0.1 < \Oh^* < 1.0$, which is well in agreement with our critical value of $\Oh^* = 0.577$. Additionally, filament profiles from Notz \& Basaran \cite{notz2004dynamics} were analysed by image analysis showing capillary wavelengths in the region of $\lambda = 2.8 R$ to $3.1R$. In fact, $\lambda = (2.96 \pm 0.15)R$ for $\Oh = 0.001$,  which is slightly lower than our inviscid value of $\lambda_{\Oh=0} = 3.63 R$. Further Ohnesorge values were explored by full 3D axisymmetric Navier-Stokes simulations using the flow solver Basilisk \cite{popinetbasilisk}. This solver uses adaptive mesh refinement and interface tracking with the volume of fluid (VoF) method. The initial configuration consisted of a square domain containing a semi-infinite free liquid cylinder of radius $R$ surrounded by a passive gas (air). The simulations were executed in a comoving frame of reference receding at the Taylor-Culick speed. Dirichlet boundary conditions for the velocity were set on the left boundary, while outlet conditions were set on the other sides of the domain. All variables were non-dimensionalised with respect to the filament radius $R$ and the liquid properties. 

We ran simulations for filaments at several values of $\Oh$ to the point where the profile behind the blob remained steady (asymptotic regime). In our analysis, for $\Oh > 1.0$, we estimated the decay length behind the blob using an exponential fitting. For $\Oh \leq 0.5$, we estimated the wavelength and decay length from the profile local maxima and minima behind the blob. These results are shown in Fig.~\ref{fig:lambda+hmin}, and show a good quantitative agreement with our model. 

\section{Conclusions\label{sec:conclusions}}
In this report, we have derived a first-order asymptotic expansion for a Newtonian liquid filament. Our analysis identifies three distinct regions: a steady filament section, an expanding spherical blob, and an intermediate \textit{matching} zone. Importantly, we have shown that, below a critical Ohnesorge number, capillary waves naturally emerge along the surface of the filament. We have analysed both their wavelength and spatial decay length along the filament as a function of $\Oh$. We analytically prove that, at the inviscid limit, the wavelength of the capillary waves is $3.63$ times the filament's radius. Our findings are found to be in agreement with numerical results in Basilisk and those corresponding to the seminal work of Notz \& Basaran \cite{notz2004dynamics}. 

\section*{Acknowledgements}

F.P.C., L.G. and J.F.M. acknowledge the financial support of Fondecyt/Iniciaci\'on No. 11170700. J.F.M. thanks the financial support of USA1899-Vridei 041931YZ-PAP Universidad de Santiago de Chile.

\bibliography{apssamp}% Produces the bibliography via BibTeX.

\end{document}